# Mechanisms of bulk and surface diffusion in metallic glasses determined from molecular dynamics simulations

Ajay Annamareddy[a)], Paul M. Voyles, John Perepezko, and Dane Morgan[b)]

Department of Materials Science and Engineering, University of Wisconsin-Madison, Madison, Wisconsin 53706, USA.

[a)] annamareddy@wisc.edu   [b)] ddmorgan@wisc.edu

The bulk and surface dynamics of $Cu_{50}Zr_{50}$ metallic glass were studied using classical molecular dynamics (MD) simulations. As the alloy undergoes cooling, it passes through liquid, supercooled, and glassy states. While bulk dynamics showed a marked slowing down prior to glass formation, with increasing activation energy, the slowdown in surface dynamics was relatively subtle. The surface exhibited a lower glass transition temperature than the bulk, and the dynamics preceding the transition were accurately described by a temperature-independent activation energy. Surface dynamics were much faster than bulk at a given temperature in the supercooled state, but surface and bulk dynamics were found to be very similar when compared at their respective glass transition temperatures. The manifestation of dynamical heterogeneity, as characterized by the non-Gaussian parameter and breakdown of the Stokes-Einstein equation, was also similar between bulk and surface for temperatures scaled by their respective glass transition temperatures. Individual atom motion was dominated by a cage and jump mechanism in the glassy state for both the bulk and surface. We utilize this cage and jump mechanisms to separate the activation energy for diffusion into two parts: (i) cage-breaking barrier ($Q_1$), associated with the rearrangement of neighboring atoms to free up space and (ii) the subsequent jump barrier ($Q_2$). It was observed that $Q_1$ dominates $Q_2$ for both bulk and surface diffusion, and the difference in activation energies for bulk and surface diffusion mainly arose from the differences in cage-breaking barrier $Q_1$.

## I. Introduction

Diffusion of atoms plays an important role in many solid-state phenomena including crystallization, formation of new phases, and aging of glasses [1]. Different parameters can be tuned to control diffusion, like changing the temperature, pressure (or density), or by adding defects. Free surfaces typically enhance diffusion as the atoms at the surface usually need to break fewer bonds to move than atoms in the bulk. An intrinsic measure of surface mobility is the in-plane surface diffusion coefficient, $D_s$. In metallic glasses (MGs), the material of interest in this work, experiments using surface-grating decay method have found that surface diffusion is much faster than bulk diffusion at the lowest temperatures studied [2–4]. Around the glass transition temperature $T_g$, $D_s \approx 10^5 \times D_b$, where $D_b$ is the bulk diffusion coefficient. Simulations have also shown the enhancement of surface diffusion over bulk in MGs [5,6]. Other glass-formers like organic molecules and polymers have shown similarly high surface enhanced diffusion in experiments [7–9]. In organic molecules, the high surface diffusion has been utilized to synthesize ultrastable glass thin films by carefully optimizing the substrate temperature and deposition rate [10–12]. The high surface mobility also recently enabled the preparation of an ultrastable metallic glass [13]. The current study investigates surface dynamics in relation to bulk dynamics and explores the enhanced surface diffusion from the viewpoint of individual motion of atoms.

Based upon MD simulations, a common understanding has been developed for the mechanisms of atomic motion inside glasses. During quenching, a glass-forming liquid exhibits a dynamical crossover in transport properties from Arrhenius to super-Arrhenius behavior [14,15]. The Arrhenius crossover temperature ($T_A$) marks the onset of heterogeneous spatially correlated dynamics. As a supercooled liquid is cooled towards $T_g$, liquid-like motion freezes and the movement of atoms is increasingly constrained by their neighbors. An atom rattles about in a cage formed by its nearest neighbors until some of them rearrange collectively to free up space large enough for the atom to jump out of the cage [16]. The need for rearrangement of neighboring atoms introduces a barrier for diffusion. The resulting slowdown in dynamics, as quantified by either the decreasing diffusion coefficient or increasing viscosity or relaxation time, is super-Arrhenius in nature and can be represented by the Vogel-Fulcher-Tammann or VFT equation [17]. The deviation from Arrhenius behavior is often described in terms of fragility, with larger deviations for more fragile glasses (strong glasses have Arrhenius behavior). Fragility is also related to the reduced Arrhenius crossover temperature ($\theta_A = T_A/T_g$), with fragile molecular liquids having $\theta_A \approx 1.4$ and strong network liquids showing $\theta_A > 2$. MGs, with intermediate fragilities, have $\theta_A \approx 2$, and $T_A$ for most metallic glasses is in the stable liquid phase above the melting temperature [14,18]. The primary ($\alpha$) relaxation in supercooled liquids approaching $T_g$ constitutes the cooperative motion of several structural units [19] and is associated with increasing energy barriers with cooling. Below $T_g$, the $\alpha$ relaxation is frozen and only short-time relaxation modes survive, characterized by localized particle hops. Here, the structural relaxation can be described by activated processes such as cage breaking or barrier hopping [20]. The primary $\alpha$ relaxation was earlier believed to completely govern the transition to the non-equilibrium glassy state. Recent experimental evidence, however, suggests that at slow cooling rates, short time-scale motions are sufficient to equilibrate systems at temperatures where $\alpha$ relaxation can be considered initially frozen [21]. Some of the fast relaxation modes are also shown to be cooperative in nature using simulations [22,23].

Compared to the bulk, the details of atomic motion on the free surfaces of metallic glasses are less studied and less well understood. For example, it is not known if the surface dynamics also become very lethargic and manifest a steep increase in activation energy while approaching some surface-related $T_g$, and if they do, how might those changes be correlated with those in the bulk? In polymer films it has been shown that the surface undergoes a glass transition at a lower temperature than the bulk [24,25]. The suppression of $T_g$ for surfaces relative to bulk is reasonable since the glass transition temperature derived from dynamics represents a crossover from (metastable) equilibrium to out-of-equilibrium, and the enhanced dynamics at the surface might be expected to reduce the temperature of this transition for the surface. The lowering of the surface $T_g$ relative to bulk has also recently been measured in a metallic glass [26].

We have investigated using simulations the suppression of surface $T_g$ in a model metallic glass. After rescaling by their respective glass transition temperatures, the surface and bulk have similar behavior and equivalent average dynamics and spatial heterogeneity in dynamics. We also explored the origin of the lower activation energy for surface versus bulk diffusion in the glassy state and propose that diffusion below $T_g$ can be understood as a combination of two activated processes: cage breaking followed by hopping. We then show that the barrier associated with cage breaking is the major contributor to the difference in bulk and surface activation energies.

## II. Simulation Methods

We used classical molecular dynamics (MD) simulations to study a $Cu_{50}Zr_{50}$ model metallic glass. The alloy was investigated over a wide temperature range, spanning liquid, supercooled, and glassy states to better understand the connection between bulk and surface dynamics. Cu-Zr alloys are the first known binary bulk metallic glasses (BMGs) and have been extensively studied using simulations [6,15,27–30].

However, the total time of MD simulations is usually less than 1 μs and extremely high cooling rates need to be employed when preparing glasses. As a result, the glasses created by MD simulations are highly under-relaxed when compared to real glasses [31,32]. This under-relaxation is reflected in the MD glass dynamics being faster by up to 10 orders of magnitude in comparison with experimental values. Nonetheless, the MD dynamics show many qualitative similarities with those seen in the experiments, including VFT behavior in the supercooled liquid and Arrhenius behavior in the glass. Therefore, consistent with many researchers in glass physics [33–35], we will assume that the MD relaxed glass yields mechanisms of bulk and surface diffusion and relative trends in surface and bulk $T_g$ similar to those found in real glasses.

The $Cu_{50}Zr_{50}$ metallic glass system was prepared as follows. A cubic simulation cell containing 16384 atoms was initially equilibrated at 2000 K for 2 ns and then quenched to 1000 K in 50 K decrements at the rate of 100 K per 6 ns. This cooling rate is sufficient to properly equilibrate the system at all temperatures from 2000 K to 1000 K as the relaxation time at the lowest temperature is ~ 1 ps. The system was subsequently cooled to 500 K in 20 K intervals at the rate of 100 K per 60 ns, corresponding to a quench rate of ~ $10^9$ K/s. The simulations were performed using the MD code LAMMPS [36]. The atomic interactions are described by the embedded atom method (EAM) potential, with the parameters taken from Mendelev *et al.* [37]. NPT conditions (at zero nominal pressure) were employed during the whole procedure and periodic boundary conditions (PBC) were used in all three directions. The temperature and pressure were controlled using the Nosè-Hoover thermostat and barostat, respectively. To integrate the classical equations of motion, the velocity-Verlet algorithm was used, with the time step set to 1.0 fs.

To simulate the bulk dynamics, the final configuration at the temperature of interest during the quenching process was used as the starting point for production runs in NVT conditions with PBC in all directions. The correct volume for the NVT simulation is obtained by using the volume from the NPT quenching. For surface analysis, with the same configuration obtained from quenching process as above, free surfaces are created by extending the simulation cell boundaries by 10 Å along the ±z-axis [30]. Again, NVT conditions and PBC are applied and the system is initially equilibrated for 1 ns to make sure the newly created surfaces at the two edges are relaxed before the production phase begins. Atoms in the outer 2.5 Å (or 12.5 Å considering the extended system dimensions) along the ±z-axis are used to evaluate the properties associated with the surface. 2.5 Å is chosen for the approximate size of monolayer as it is the size of a nearest neighbor shell determined from the first maximum of the bulk radial distribution function of atomic density. The same simulations can also be used to measure bulk dynamics by choosing atoms in a finite region symmetric around the center. However, the 3D periodic systems exhibit better convergence. It is to be noted that creation of free surfaces can introduce compositional gradients from the surface towards the bulk due to surface segregation. In Cu-Zr alloys, Cu atoms segregate to the surface, likely because they have a smaller coordination number than Zr and so fewer bonds needed to be broken for Cu relative to Zr at the surface. In this work, we observed that the MG composition changes from $Cu_{50}Zr_{50}$ in the bulk to $Cu_{54}Zr_{46}$ at the surface (composition measured in the outer 2.5 Å layer). In Supporting Information Section S1, we show that the mobility of Cu and Zr atoms are very similar in $Cu_{50}Zr_{50}$ and $Cu_{54}Zr_{46}$ and, hence, the change in composition at the surface will not have a significant effect on the surface dynamics results presented in this work. Free surfaces creation also generates significant strains on atoms at the boundary leading to strain-relaxation and this is different from the diffusion motion that we are interested in this study. We will show in the Supporting Information Section S2 that after the 1 ns equilibration period for surface simulations, the relaxation effects are quite insignificant.

The diffusion coefficient is calculated based on the mean-squared displacement (MSD) of atoms as:

$$D = \lim_{t \to \infty} \frac{1}{(2d)Nt} \left\langle \sum_{i=1}^{N} |r_i(t) - r_i(0)|^2 \right\rangle \quad (1)$$

where $d$ is the number of dimensions, $N$ is the number of atoms, $r_i(t)$ is the position of atom $i$ at time $t$, and $\langle \rangle$ refers to the ensemble average. For the bulk diffusion coefficient ($D_b$), $d = 3$. As free surfaces inhibit the motion normal to a plane, in the case of surface diffusion ($D_s$), only the lateral displacement of atoms in the plane is considered and $d = 2$. To evaluate $D_s$, the subset of atoms contributing to Eq. (1) can be identified in different ways, considering (a) atoms restricted to the surface layer continuously from $t = 0$ to $t = t$, (b) atoms belonging to the surface layer at $t = 0$, or (c) atoms belonging to surface layer at both $t = 0$ and $t = t$. Method (a) inherently captures atoms that are, on average, less mobile and the sample size is greatly reduced at longer times, leading to both biased and deteriorating statistics. Methods (b) and (c) give similar results, although we find that method (c) leads to better convergence and is therefore adopted in this work. To further improve statistics, the method of buffer averaging as described in [38] is employed. Relaxation time ($\tau_\alpha$) is extracted from self-intermediate scattering function, defined as:

$$F_s(q,t) = \frac{1}{N} \left\langle \sum_{n=1}^{N} e^{i\mathbf{q} \cdot [\mathbf{r}_n(t) - \mathbf{r}_n(0)]} \right\rangle$$

and $F_s(q, \tau_\alpha) = 1/e$. The chosen wave vector, $|\mathbf{q}| = 2.7$ Å$^{-1}$ in Cu$_{50}$Zr$_{50}$, coincides with the maximum of the structure factor. For the surface $F_s$, $\mathbf{q}$ is taken parallel to the surface.

### III. Results

#### 1. *Diffusion at bulk and surface*

From the temperature dependence of volume and enthalpy, the bulk glass transition temperature ($T_{g,b}$) of Cu$_{50}$Zr$_{50}$ is estimated to be 700 K ± 10 K, which is higher than the reported experimental value of 666 K [39]. We expect a somewhat higher glass transition temperature due to an orders of magnitude faster cooling rate in our simulation when compared to experiments, although some errors from the EAM potential are also expected. In Cu$_{50}$Zr$_{50}$ alloy, Cu atoms are more mobile than Zr [15] and we will only focus on the dynamics of the entire system here. For completeness, Fig. S5 in the Supporting Information shows the variation of $D_{Cu}$, $D_{Zr}$, and $D_{Cu}/D_{Zr}$ with temperature. Fig. 1 shows the variation of bulk ($D_b$) and surface ($D_s$) diffusion coefficients of Cu$_{50}$Zr$_{50}$ as a function of temperature in an Arrhenius plot. At high temperatures (above 1000 K), $D_b$ and $D_s$ are almost identical. Fitting the bulk liquid dynamics to the Arrhenius equation $D = D_0 e^{-Q/(k_B T)}$ yields $Q = 0.432 \pm 0.007$ eV, where the uncertainty represents one standard deviation error of the error derived from the linear fitting on the log plot. The same methods are used, and one standard deviation given, for all activation energy and log of pre-exponential factor ($\ln(D_0)$) values in this paper. $D_0$ error bars are obtained from $\ln(D_0)$ by error-propagation methods. More details on $D_0$ and $\ln(D_0)$ errors are given below. With further cooling, at ~1000 K, the slowdown of the bulk dynamics becomes rapid with a shift from Arrhenius to super-Arrhenius behavior. The VFT equation $D = D_0 e^{-Q/(k_B(T-T_0))}$ represents the variation of $D_b$ in this regime quite well. The equation implies that the relaxation time of the system diverges (or $D \to 0$) at a finite temperature $T_0$ below $T_{g,b}$, although experiments have not found compelling evidence of this behavior [40] and the equation is considered mostly empirical in nature. In the glassy regime, starting at $T_{g,b}$, bulk diffusion can again be represented by the Arrhenius equation even though the range of temperatures studied is rather limited because of the extremely slow bulk dynamics. The fitting yields an activation energy ($Q$) = 1.66 ± 0.07 eV and $D_0$ = 1.49 ± 2.42 m$^2$/s. It is to be noted that while $Q$ is normally distributed, the distribution of $D_0$ is *log*-normal (i.e., ln ($D_0$) is normally distributed). Therefore, the standard error in $D_0$ can be misleading and negative values of $D_0$ are not possible despite the standard error being larger than the mean value. To provide a more useful error estimate we also provide values of ln ($D_0$) and their errors for all cases in Table 1. Experimental diffusion calculations have also indicated the VFT nature of supercooled dynamics and the Arrhenius nature of glassy dynamics in many metallic glasses [41], including a recent study on $Cu_{50}Zr_{50}$ [42].

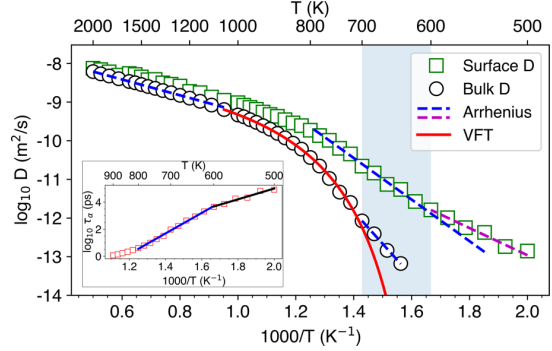

FIG. 1. Bulk and surface diffusion in $Cu_{50}Zr_{50}$ as a function of temperature. The shaded region indicates the range of temperature when the bulk transforms into a glass while the surface is still in the supercooled liquid state. For bulk in the supercooled state, the VFT equation holds well over a large temperature range and the fitting parameter $T_0$ is 572 ± 10 K. (Inset) Arrhenius plot of relaxation time of surface atoms. Similar to surface diffusion, a kink is observed at 600 K, which is recognized as the surface glass transition temperature ($T_{g,s}$).

TABLE 1: Values of the fitting parameters for the Arrhenius and VFT equations applied on $D$, $\Gamma$, and $<l^2>$.

|  | *State/temperature-range* | ln (pre-factor) | Pre-factor | $Q$ |
|---|---|---|---|---|
| Bulk $D$§ | Liquid (2000 K – 1000 K) | -16.36 ± 0.06 m$^2$/s | 7.9 × 10$^{-8}$ ± 4.7 × 10$^{-9}$ m$^2$/s | 0.432 ± 0.007 eV |
|  | SC liquid (1000 K – 700 K)$^\$$ | - | 8.2 × 10$^{-9}$ ± 7.6 × 10$^{-10}$ m$^2$/s | 2.13 ± 0.16 eV |
|  | Glass (700 K – 640 K) | -0.25 ± 1.14 m$^2$/s | 1.49 ± 2.42 m$^2$/s | 1.66 ± 0.07 eV |
| Bulk $Q_1$† | Glass (700 K – 640 K) | 18.1 ± 0.6 /ns | 8.4 × 10$^7$ ± 5.7 × 10$^7$ /ns | 1.28 ± 0.04 eV |
| Bulk $Q_2$‡ | Glass (700 K – 640 K) | 4.1 ± 0.1 Å$^2$ | 60 ± 7 Å$^2$ | 0.120 ± 0.006 eV |
| Surface $D$§ | Liquid (2000 K – 1000 K) | -16.48 ± 0.12 m$^2$/s | 7.0 × 10$^{-8}$ ± 8.5 × 10$^{-9}$ m$^2$/s | 0.365 ± 0.014 eV |
|  | 700 K – 600 K | -8.75 ± 0.72 m$^2$/s | 2.0 × 10$^{-4}$ ± 1.7 × 10$^{-4}$ m$^2$/s | 0.95 ± 0.04 eV |
|  | Glass (600 K – 500 K) | -14.57 ± 0.93 m$^2$/s | 7.3 × 10$^{-7}$ ± 8.5 × 10$^{-7}$ m$^2$/s | 0.65 ± 0.04 eV |
| Surface $Q_1$† | 700 K – 640 K | 11.39 ± 0.65 /ns | 1.1 × 10$^5$ ± 7.8 × 10$^4$ /ns | 0.70 ± 0.04 eV |
| Surface $Q_2$‡ | 700 K – 640 K | 6.40 ± 0.34 Å$^2$ | 638 ± 222 Å$^2$ | 0.23 ± 0.02 eV |

§: Arrhenius fitting shown in Fig. 1; †: Fig. 5(b); ‡: Fig. 5(c)
$: VFT fitting shown in Fig. 1 and $T_0$ = 572 ± 10 K.

$D_s$ closely follows $D_b$ in the liquid state, above 1000 K. As $D_b$ exhibits a marked slowdown at intermediate temperatures and the non-Arrhenius slowdown of $D_s$ is relatively weak, ($D_s/D_b$) increases with cooling. Below $T_{g,b}$, $D_s$ can also be described by the Arrhenius equation. In the temperature range from 700 K to 600 K, $Q$ = 0.95 ± 0.04 eV and $D_0$ = 2.0 × 10$^{-4}$ ± 1.7 × 10$^{-4}$ m$^2$/s. The ratio of activation energy of



surface to bulk diffusion in this regime is 0.57, which is close to 0.5 that has been observed in earlier glass measurements [7]. The value of $Q_s$ being half the value of $Q_b$ is generally attributed to the reduction in the number of bonds surface atoms need to break compared to bulk atoms for diffusion. However, the difference in the values of $D_0$ for bulk and surface pre-exponential factors (by 4 orders of magnitude in our case) is also significant. The origin of $D_0$ differences between bulk and surface is not clear and further work is needed. However, we do not try to address this question further in this paper. Several experimental studies on bulk glass diffusion have utilized the analytical expression originally derived for crystals to analyze $D_0$ [43–46], but the atomistic contributions controlling its value are still not clear in glasses.

The high mobility of surface atoms allowed the calculation of $D_s$ at much lower temperatures than $D_b$. $D_s$ varies smoothly across $T_{g,b}$ and shows a kink at 600 K ± 10 K, about 100K below $T_{g,b}$. We associate this kink with the surface glass transition temperature ($T_{g,s}$). Below $T_{g,s}$, $D_s$ again varies in an Arrhenius manner with $Q = 0.65 \pm 0.04$ eV and $D_0 = 7.3 \times 10^{-7} \pm 8.5 \times 10^{-7}$ m$^2$/s. The values of the fitting parameters are also given in Table 1. A similar change in slope is also observed for the structural relaxation time of the surface at 600 K ± 10 K, shown as the inset of Fig. 1. For a Cu composition consistent with the surface segregation, namely $Cu_{54}Zr_{46}$, $T_g$ is identified to be 720 K as shown in Fig. S1(a). This is an increase of approximately 20 K compared with $T_g$ of $Cu_{50}Zr_{50}$. Assuming the surface and bulk composition trends are qualitatively similar, this result shows that the lowering of $T_{g,s}$ for the Cu-rich segregated surface compared to the bulk glass transition is not caused by segregation but a consequence of fast surface dynamics. Segregation, does introduce some small changes in our measurement of $T_{g,s}$ for $Cu_{50}Zr_{50}$, but that the effect goes in the direction of increasing $T_{g,s}$ relative to an unsegregated surface. We also note that increasing the thickness of the surface (currently 2.5 Å) will increase the observed $T_{g,s}$ towards $T_{g,b}$. Below, we will compare the dynamics of the bulk and surface when viewed in terms of $T_{g,b}$ and $T_{g,s}$, respectively.

## 2. Comparing bulk and surface dynamics

Figs. 2(a) and 2(b) show the diffusion coefficients and relaxation times of the bulk and surface with the temperatures normalized to their respective glass transition temperatures. The figures show that upon approaching the transition, the bulk and surface dynamics are quite similar. The very disparate values of $D_b$ and $D_s$ in the supercooled state in Fig. 1 are therefore primarily a reflection of the underlying differences in the degree of supercooling. However, the curvatures of bulk and surface dynamics when approaching $T_g$ are not identical. Taking the case of relaxation time, the steeper increase in bulk $\tau_\alpha$ around $T_g$ indicates that the bulk is more fragile than the surface. The larger bulk fragility indicates that the bulk system occupies a more rugged part of the potential energy landscape and needs to overcome large barriers for relaxation [47]. The impact can be seen in a faster growth of bulk relaxation time with cooling compared to surface relaxation below $T_g$. Below the glass transition, a clear distinction is evident between the bulk and surface – the bulk dynamics slows down faster than surface dynamics.

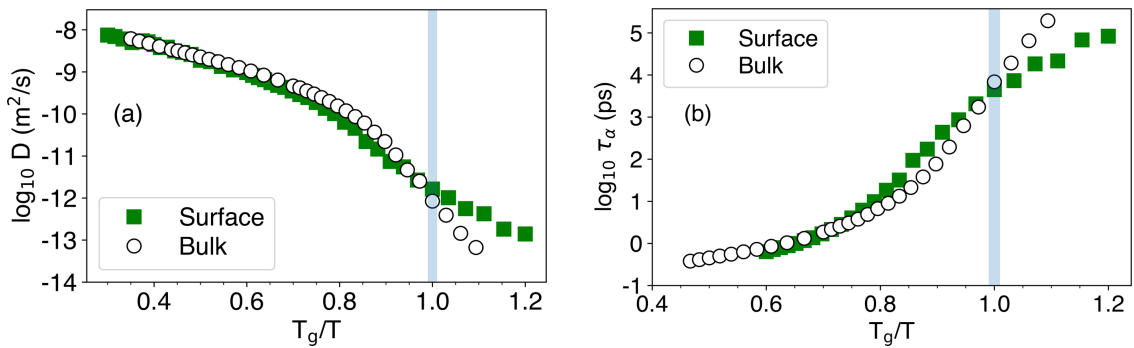



FIG. 2. Similarities in bulk and surface dynamics when examined as a function of $T_g/T$ for $T_g = T_{g,b}$ and $T_{g,s}$, respectively. Arrhenius plots of (a) diffusion coefficients and (b) relaxation times of bulk and surface atoms as a function of normalized temperature. Blue bars indicate $T = T_g$.

An important aspect of the slowdown dynamics associated with glasses is dynamic heterogeneity (DH), which refers to the broadening of the distribution of particle mobilities or relaxation times in the system along with the spatial clustering of particles with similar mobilities [48]. Even though Fig. 2 demonstrates that the average dynamics are similar between bulk and surface in terms of normalized glass transition temperatures, the corresponding heterogeneities could be very different. To investigate the heterogeneities, we measure the non-Gaussian parameter ($\alpha_2$) which is a standard metric to quantify mobility fluctuations in supercooled liquids and glasses [49] and is defined as:

$$\alpha_2(\Delta t) \equiv \frac{3\langle r^4(\Delta t)\rangle}{5\langle r^2(\Delta t)\rangle^2} - 1. \tag{2}$$

Here, $r(\Delta t)$ refers to the distance traveled by an atom from its position at $t = 0$ after time $\Delta t$. The $\alpha_2$ parameter characterizes the deviation of particle mobility from a Gaussian distribution and is defined so that for Brownian motion, it is equal to zero. Figure 3(a) shows the temporal variation of $\alpha_2$ for bulk and surface atoms at their $T_g$ (700 K and 600 K, respectively), with time scales normalized to the corresponding relaxation times. The peak values of $\alpha_2$, quantifying the degree of heterogeneity in mobility, are very close, indicating similar heterogeneity for bulk and surface.

Another consequence of the broad distribution of particle mobility in the supercooled state is the decoupling between viscosity and diffusivity or the breakdown of Stokes-Einstein (SE) relation, which results from the different ways transport and relaxation properties sample the distribution. The SE equation relates diffusion ($D$) and reduced shear viscosity ($\eta \equiv \eta_s/T$, where $\eta_s$ is the shear viscosity) as $D\eta =$ constant. In supercooled liquids, the structural relaxation time ($\tau_\alpha$) is proportional to $\eta$ so that the SE relation can be rewritten as $D\tau_\alpha =$ constant [50]. Fig. 3(b) shows the variation of $D\tau_\alpha$ for bulk and surface as a function of temperature normalized to their respective $T_g$'s. We notice that $D\tau_\alpha \sim$ constant in the high temperature liquid regime above 1000 K for the bulk. For the surface atoms, Fig. 1 shows that slowdown in diffusion starts to manifest around 800 K and in Fig. 3(b), $D\tau_\alpha \sim$ constant above this temperature. $D\tau_\alpha$ starts to rise in the supercooled regime with nearly identical variation at all lower temperatures for bulk and surface. Based on the similarities observed in Figs. 2 and 3, for this system it seems that a useful way to connect bulk and surface dynamics is to measure values relative to their different glass transition temperatures, although further study is needed to see if this behavior is exhibited more generally.

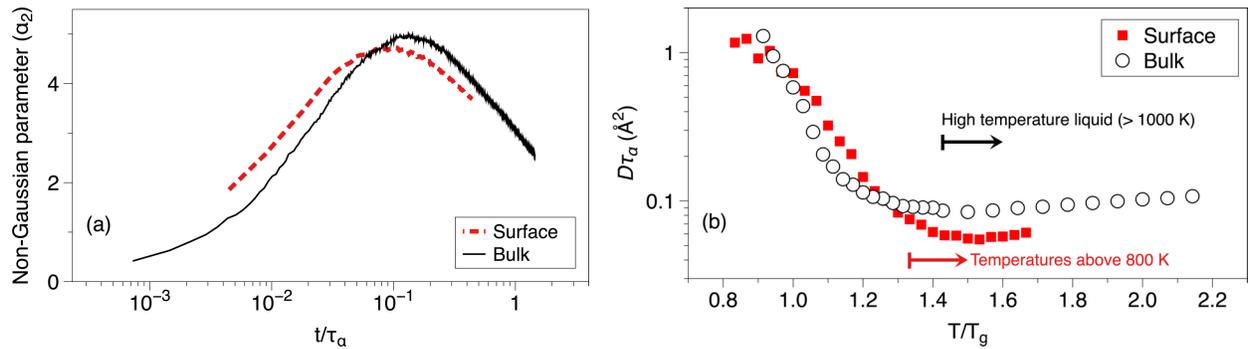



FIG. 3. Demonstration of similar heterogeneities for surface and bulk in a metallic glass. (a) Temporal dependence of non-Gaussian parameter for the bulk and surface at their respective $T_g$. (b) Break-down of SE relation as quantified by the variation of $D\tau_\alpha$ for bulk and surface. The temperature regimes identified by black and red arrows correspond to bulk and surface, respectively.

### *3. Diffusion mechanisms and origins of activation energies*

An approach that has been used to study the slowing down in glasses in both experiments and simulations is to focus on the individual- or single-atom dynamics [28,51–53]. This method is relevant in the supercooled and glassy states where cage-jump motion becomes increasingly pronounced. Our motivation here is to apply this approach to understand the differences in diffusion activation energy for the bulk and surface in glasses.

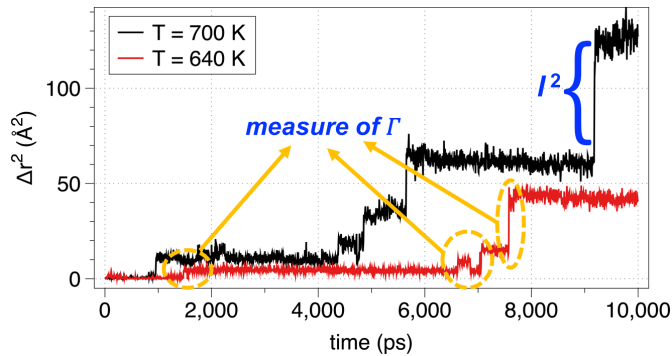

FIG. 4. Typical squared displacement of a mobile atom at 700 K and 640 K in the bulk. The trajectories indicate long periods of caging punctuated by sudden jumps.

Fig. 4 shows the typical squared displacements of a mobile atom in the bulk at 700 K (= bulk $T_g$) and 640 K. The atoms in the bulk are mostly caged, and the atomic displacements are sudden (< 1 ps) with substantial rearrangements. According to Lü and Wang [28], this kind of cage-jump motion is noticeable at temperatures up to 860 K in $Cu_{50}Zr_{50}$. The localized particle rearrangements are β-relaxation events carrying the system between neighboring basins on the energy landscape as described by Stillinger [47]. Surface atoms also behave similarly, although the fraction of time they are caged is smaller than bulk atoms. Keeping this in mind, starting at bulk $T_g$, we model diffusion as a series of discrete jumps of individual atoms and, analogous to in crystals, write diffusion as [54]

$$D = \left( \Gamma \frac{\langle l^2 \rangle}{2d} \right) f \qquad (3)$$

where $\Gamma$ is the jump rate, $f$ is the correlation factor, $\langle l^2 \rangle$ is the average of squared jump length, and $d$ is the number of dimensions. The term in brackets is derived by assuming that atom jumps follow ideal random-walk during diffusion, and $f$ (which is less than 1) accounts for the non-randomness or back-and-forth correlated jumps. In the following, we approximate $f$ to 1 and remove correlated jumps in the calculation of $\Gamma$ and $\langle l^2 \rangle$. Eq. (3) ignores the diffusion accrued by small-scale rearrangements and should be more accurate at low temperatures in the glassy state, where diffusion is dominated by jump motion of atoms.



In order to evaluate $\Gamma$ and $\langle l^2 \rangle$, the approach followed by Lü and Wang [28] is adopted. To detect a jump, the positions over a time window are averaged to remove the effect of localized vibrations. The refined or time-coarse-grained position of particle $i$, $\bar{\mathbf{r}}_i(t)$, over a time window $\Delta t$ is defined as:

$$\bar{\mathbf{r}}_i(t) = \frac{1}{\Delta t} \int_0^{\Delta t} \mathbf{r}_i(t+t')dt'. \tag{4}$$

Following [28], $\Delta t = 200$ ps is applied, and the trajectories are separated in time by 5 ps. A jump is identified if the standard variance of the time-coarse-grained position is larger than $5\langle u^2 \rangle$, where $\langle u^2 \rangle$ is the temperature-dependent Debye-Waller factor of the system. $\langle u^2 \rangle$ is defined as the MSD corresponding to the inflection point in the log [MSD] – log [time] curve and characterizes the length scale associated with cage rattling. To identify jumps that start at the edges of a time window, the trajectories are actually analyzed with the time windows shifted by $\Delta t/2$ [55]. For a jump identified in time window $j$ for atom $i$, the jump vector is defined as: $\mathbf{l}_i^j = \left| \bar{\mathbf{r}}_i^{j+1} - \bar{\mathbf{r}}_i^{j-1} \right|$, where $\bar{\mathbf{r}}_i^{j+1}$ and $\bar{\mathbf{r}}_i^{j-1}$ are the coarse-grained positions of a jumping particle $i$ in time windows $j + 1$ and $j – 1$, respectively. The jump length, $l = |\mathbf{l}|$. If an atom makes a jump in two or more successive time windows, the whole motion is regarded as a single jump that started in the first and ended in the last time window. By having the jump length be at least $\sqrt{\langle u^2 \rangle}$, we remove both forward-backward correlated jumps and loop motion of atoms. Forward-backward jumps of an atom separated in time are also eliminated if for successive jumps $\left| \mathbf{l}_i^{n+1} - \mathbf{l}_i^n \right| < 1.0$ Å. Using this methodology, we evaluated $\Gamma$ and $\langle l^2 \rangle$ for bulk and surface atoms, leading to $D$ using Eq. (3).

The jump rate ($\Gamma$) slows with cooling, and we will show that it follows Arrhenius behavior in the glassy regime. We speculate that at low temperatures, the jump of an atom necessitates overcoming a barrier associated with the creation of free space by the collective rearrangement of neighboring atoms. Hence, we equate the activation energy ($Q_1$) obtained from the Arrhenius variation of $\Gamma$ to the barrier associated with the rearrangement of neighboring atoms. We call this the *cage-breaking barrier* as the cage is no longer rigid and the neighboring atoms facilitate an escape route. $\langle l^2 \rangle$ will also be shown to follow the Arrhenius relation, and we call the associated activation energy ($Q_2$) the *hopping barrier*. In this picture, an atom that breaks its cage and makes a jump has a finite time before getting caged again. If, during this interval, it can jump again, it only has to overcome the hopping barrier ($Q_2$). We therefore combine the jumps that are close together in time and treat them as a single jump during the calculation of $\Gamma$. The variation of $\langle l^2 \rangle$ mainly stems from the frequency of jumps spaced closely in time and is dictated by $Q_2$. The total activation energy of diffusion ($Q_D$) is $Q_D = Q_1 + Q_2$. In writing $Q_D$ as a simple summation, we assume that the two processes are independent of each other. This separation of $Q_D$ closely follows the separation of $Q_D$ in crystals as the sum of formation and migration activation energies. We can think of $Q_1$ as the energy to form the defect that mediates motion, here a broken cage, similar to a vacancy in a crystal.

Fig 5(a) shows that $D$ computed from the jump data and Eq. 3 is in good agreement with $D$ obtained from Eq. (1). Figs. 5(b) and 5(c) show Arrhenius plots of $\Gamma$ and $\langle l^2 \rangle$, with the corresponding activation energies. The average jump length in an individual time window is very weakly dependent on temperature and increases linearly from 640 K (2.12 Å) to 700 K (2.15 Å). Hence, the variation of $\langle l^2 \rangle$ reflects the increasing number of jumps in successive time windows with increasing temperature. Fig. 6 compares the surface and bulk activation energies in the glassy regime. The main contribution to the higher activation energy for bulk diffusion compared to surface diffusion is the cage-breaking barrier, $Q_1$. In the Supporting Information Section S4, it is demonstrated that even for surfaces of larger thickness, the surface activation energies are very similar to the $Q$'s shown in Fig. 6. Hence, the difference in $Q_1$ for the bulk and surface accounts for almost all of the total activation energy ($Q_D$) differences between bulk and surface. In Section S5 of the Supporting Information, we vary the time-window $\Delta t$ by almost two orders of magnitude to study the effect of $\Delta t$ on $Q$'s. This analysis shows that our choice of $\Delta t = 200$ ps is very reasonable for studying single atom dynamics and a proper choice of $\Delta t$ is important to determine accurate values of $Q_1$ and $Q_2$.

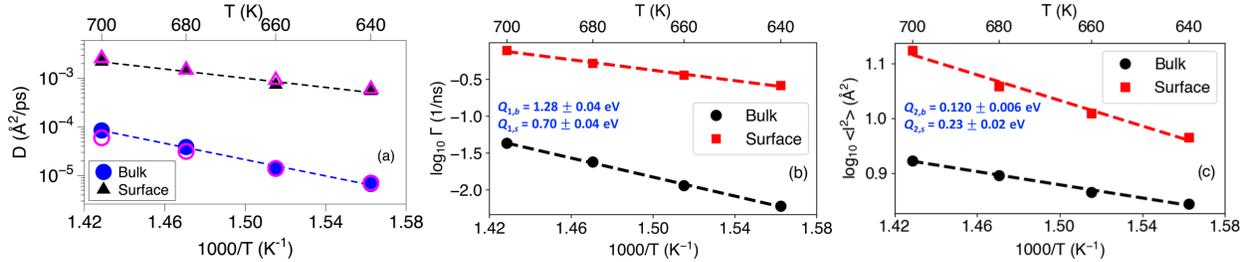

FIG. 5. (a) Comparison of diffusion coefficients obtained from the mean-squared displacement of atoms (solid symbols), as given by Eq. (1), and the discrete-jump model (open symbols), from Eq. (3). Arrhenius plots of (b) $\Gamma$ and (c) $\langle l^2 \rangle$ for surface and bulk atoms.

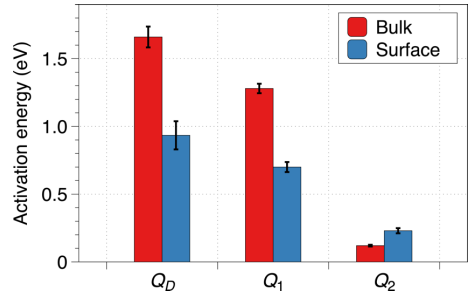

FIG. 6. Comparison of activation energies for total diffusion ($Q_D$), cage-breaking ($Q_1$), and hopping ($Q_2$) for surface and bulk atoms.

The migration barriers ($Q_2$) for the bulk and surface are quite similar, differing by ~ 0.1 eV. The accuracy of $Q_2$ from Fig 5(c) can be verified by using the nudged elastic band (NEB) method on atom hops observed in the MD simulations. First, the inherent structures are obtained by energy minimization of MD configurations separated by 5 ps. Atoms that are displaced more than 1.8 Å, the first minimum in the van Hove self-correlation function [23] of $Cu_{50}Zr_{50}$, between successive inherent structures are identified as jump events. Then, the NEB method is applied to the position of all atoms before the jump and the position of atoms in a sphere of radius 5 Å centered on the jumping atom after the jump to find the migration barrier energies. We performed the analysis on 1000 randomly chosen jump events each from the bulk and the surface at 640 K, the lowest bulk temperature studied. The low temperature minimizes the displacement of atoms not connected to a jump event and should therefore give the best estimate of the barrier. A histogram of the barrier energies is shown in Fig. 7. For both bulk and surface, the barriers are spread over a range of energies reflecting the diverse local atomic arrangements seen by jumping atoms. To derive average migration barrier ($\bar{E}$), we perform a parallel rate analysis, assuming that the jumping atoms have access to multiple barriers simultaneously consistent with multiply-connected nature of the potential energy landscape for glasses [56]. This assumption corresponds to calculating the average jump rate as $\langle e^{-\beta E}\rangle$ over all the measured barrier energies and, $e^{-\beta \bar{E}} = \langle e^{-\beta E}\rangle$. The resulting hopping barriers are $0.17 \pm 0.19$ eV and $0.18 \pm 0.21$ eV for bulk and surface, respectively, and compare well with the 0.12 eV and 0.23 eV obtained by using the cage-jump model.

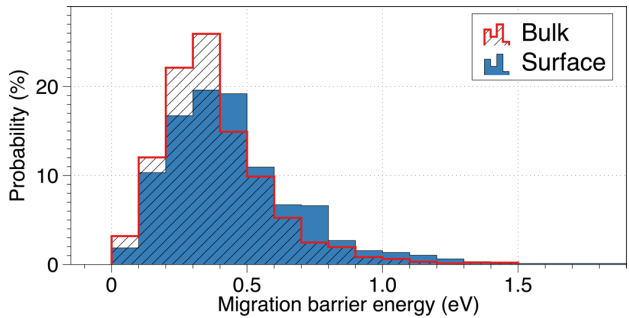

FIG. 7. Histogram of bulk and surface hopping barrier energies corresponding to atom jumps recognized in MD simulations at T = 640 K. The barriers are measured using NEB method.

Finally, we want to address the issue of applying the cage-jump model to surface atoms. Surface atoms have ample free space above the surface and enhanced lateral surface diffusion may have a dominant contribution from atoms diffusing into this open space. Perhaps the most obvious mechanism to use this space is to diffuse as adatoms (which are on top of all their neighbors) in the plane. In the case of adatom motion, the cage-jump model may not be applicable as the atom is not caged in the direction of motion. Alternatively, enhanced surface diffusion can be enabled by easy motion of atoms in the plane, in which case the cage-jump model is applicable. In thin polymer films, the surface is considered as almost liquid-like, with many orders of magnitude higher mobilities compared to the bulk near the $T_g$ [25,57], which is consistent with diffusion inside the surface layer, rather than as adatoms. We can identify adatoms as atoms with low coordination number (CN ≤ 6 for Cu and CN ≤ 8 for Zr in this system) that sit on or near the surface of the model. Fig. 8 shows the layering of density close to the surface (left y-axis) and the average CN of Cu and Zr atoms near the surface (right y-axis). Values less than 0 on the x-axis in the figure indicate the free space available for surface atoms. Both CNs fall below the adatom threshold only for $z < -0.4$ Å, where the normalized atom density is ≤ 10%. Fig. 9 shows the contour map of coordination numbers before and after the jump for all the surface atom jumps, including both adatoms and the atoms in the first monolayer from $z = 0$ to 2.5 Å, for Cu and Zr atoms separately. Regions in the contour corresponding to adatom hops are bounded by dotted lines. The contour plots show that most surface atom jumps in our simulations do not correspond to adatom diffusion and so our hypothesis that most surface atoms need to break their cages in order to diffuse seems to be valid.

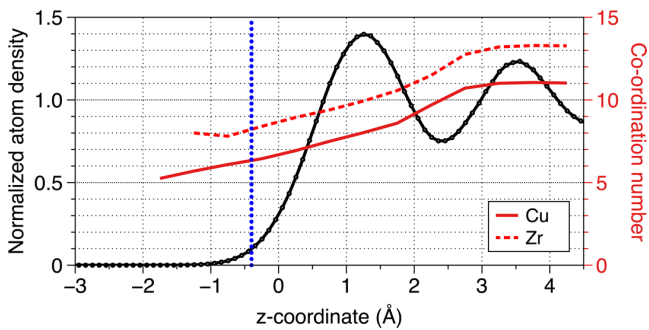

FIG. 8. Profile view of the normalized atom density and co-ordination number of Cu and Zr atoms in $Cu_{50}Zr_{50}$ at T = 700 K. The vertical blue line identifies the z-coordinate threshold for atoms to be considered adatoms due to their low coordination number.

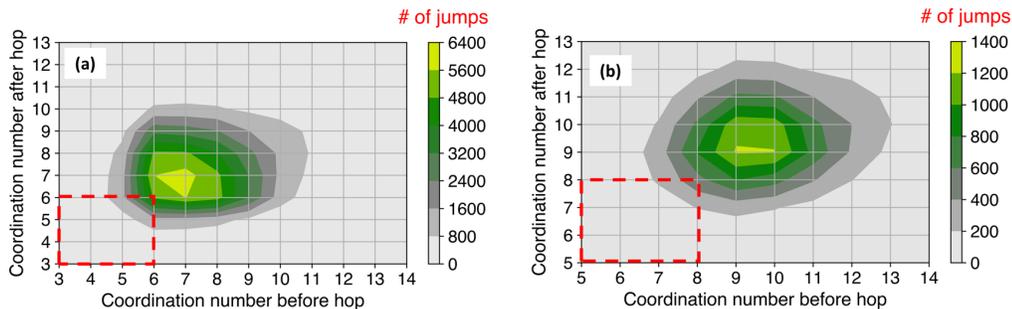

FIG. 9. Contour view of the distribution of surface atom jumps in terms of co-ordination number of the atom before and after the jump for (a) Cu and (b) Zr atoms. Dotted regions indicate the contribution arising from adatom-like jumps.

The surface cage breaking barrier $Q_{1,s}$ is about half the bulk cage breaking barrier, $Q_{1,b}$. This difference accounts for most of the difference in the total diffusion barrier energy, $Q_D$. We attribute the lower $Q_1$ to the reduced number of atoms in the cage for the surface atoms. Essentially, the top half of the bulk cage is missing on the surface, so the number of atoms that must rearrange is significantly reduced.

In this work, we did not provide any atomistic picture of cage-breaking mechanism, and we do not think that such a picture is easy to obtain. Cage breaking in metallic glasses is very likely cooperative [23], involving many atoms, and therefore requires a complex description. Furthermore, many authors have sought for such an understanding without obtaining it, as illustrated by the partial solutions presently available. For example, Ma and collaborators have shown that slow dynamics in glasses correlates with larger fractions of icosahedral local neighbor shells, but no quantitative picture of the cage structure controlling dynamics and its breaking mechanism have been proposed [58]. Perhaps the most successful quantitative effort in this direction comes from Liu and collaborators. They have been able to establish a feature of local environments, called softness, that correlates with atomic motion. However, their softness parameter was derived by applying machine learning to a collection of 166 local structure functions [59]. Their work demonstrates that the state of the art for descriptions of the local environments that control hopping are still far from a simple human-understandable atomic description.

The results in this manuscript provide both physical insight into the process of surface diffusion, based on caging and hopping, and an easy path to estimate surface mobility by rescaling more readily accessible bulk data by $T_{g,s}$. The temperature difference between $T_{g,s}$ and $T_{g,b}$ probably depends both on cooling rate and materials system, and reports range from 20 K for metals [26] to around 70 K in freestanding polymer films [25] for experimental cooling rates, to the 100 K reported here for MD cooling rates. Greater quantitative understanding of surface dynamics would aid the synthesis of metallic glasses in thin film form, including synthesis of glasses with enhanced stability [13], and synthesis of nanostructures via surface-mediated molding methods [60]. It may also help explain the surface crystallization of metallic glasses near $T_g$ [61] and even anomalous friction behavior near $T_g$ [62].

## IV. Conclusions

We have studied the bulk and surface dynamics of a model binary metallic glass-former using molecular dynamics simulations. As the alloy is supercooled, the bulk dynamics exhibits a rapid slowdown, as represented by the VFT equation, before undergoing a glass transition at $T_{g,b}$. In contrast, the surface dynamics manifests a more gradual slowdown and exhibits a glass transition at $T_{g,s}$, 100 K below $T_{g,b}$. We also demonstrated an equivalence of average dynamics and dynamical heterogeneity between bulk and surface, when compared at temperatures normalized by the appropriate $T_g$. However, the fragilities of the bulk and surface liquids are not the same, so the behavior is not completely identical. The agreement between bulk and surface relaxation times at their respective $T_g$ (even though relaxation times are not used in the determination of either $T_g$) illustrates the dynamical nature of the glass transition for *both* bulk and surface.

Single atom dynamics in the glassy state indicates that the dynamics in the bulk and on the surface are both overwhelmingly dominated by cage and jump motion. The dynamics of bulk and surface dynamics exhibit Arrhenius behavior below bulk $T_{g,b}$. In both cases, the energy barriers for individual atom hops are much smaller than their respective Arrhenius diffusion activation energies. The Arrhenius nature of the temperature-dependent cage-breaking rate suggests that the breaking of the cage involves a barrier, in addition to the hop barrier described above, that contributes to the overall diffusion activation energy. Hence, by treating diffusion as a series of discrete jumps of atoms, as in crystals, the resulting analytical expression for diffusion separates the activation energy for diffusion ($Q_D$) into two parts: (i) cage-breaking barrier ($Q_1$), and (ii) the subsequent jump barrier ($Q_2$). We find that $Q_1$ dominates $Q_2$ for both bulk and surface diffusion, and $Q_1$ of the bulk is almost twice the $Q_1$ of the surface. We attribute this to the reduced dimensionality of the surface in which cage-breaking requires the rearrangement of fewer atoms (around half) compared to the bulk.

**Acknowledgements**

The authors are grateful to the Extreme Science and Engineering Discovery Environment (XSEDE), which is supported by National Science Foundation grant number OCI-1053575, and the Center for High Throughput Computing at UW-Madison for the computing resources. This work was supported by the University of Wisconsin Materials Research Science and Engineering Center (DMR-1720415).

**References:**

[1]   P. Heitjans, J. Kärger, Diffusion in Condensed Matter, Springer, 2005.

[2]   C.R. Cao, Y.M. Lu, H.Y. Bai, W.H. Wang, High surface mobility and fast surface enhanced crystallization of metallic glass, Appl. Phys. Lett. 107 (2015) 141606.

[3]   C.R. Cao, L. Yu, J.H. Perepezko, Surface dynamics measurement on a gold based metallic glass, Appl. Phys. Lett. 116 (2020) 231601.

[4]   K.L. Ngai, S. Capaccioli, C.R. Cao, H.Y. Bai, W.H. Wang, Quantitative explanation of the enhancement of surface mobility of the metallic glass $Pd_{40}Cu_{30}Ni_{10}P_{20}$ by the Coupling Model, J. Non. Cryst. Solids. 463 (2017) 85–89.

[5]   B. Böddeker, H. Teichler, Dynamics near free surfaces of molecular dynamics simulated $Ni_{0.5}Zr_{0.5}$ metallic glass films, Phys. Rev. E. 59 (1999) 1948–1956.

[6]   H. Chen, B. Qu, D. Li, R. Zhou, B. Zhang, Atomic structure and dynamics properties of $Cu_{50}Zr_{50}$ films, J. Appl. Phys. 123 (2018) 025307.

[7]   L. Zhu, C.W. Brian, S.F. Swallen, P.T. Straus, M.D. Ediger, L. Yu, Surface self-diffusion of an organic glass, Phys. Rev. Lett. 106 (2011) 1–4.

[8]   W. Zhang, L. Yu, Surface diffusion of polymer glasses, Macromolecules. 49 (2016) 731–735.

[9]   Y. Chen, W. Zhang, L. Yu, Hydrogen bonding slows down surface diffusion of molecular glasses, J. Phys. Chem. B. 120 (2016) 8007–8015.

[10]  S.F. Swallen, K.L. Kearns, M.K. Mapes, Y.S. Kim, R.J. McMahon, M.D. Ediger, T. Wu, L. Yu, S. Satija, Organic glasses with exceptional thermodynamic and kinetic stability, Science. 315 (2007) 353–356.

[11]  M.D. Ediger, Perspective: Highly stable vapor-deposited glasses, J. Chem. Phys. 147 (2017) 210901.

[12]  L. Berthier, P. Charbonneau, E. Flenner, F. Zamponi, Origin of ultrastability in vapor-deposited glasses, Phys. Rev. Lett. 119 (2017) 1–5.

[13]  H.-B. Yu, Y. Luo, K. Samwer, Ultrastable metallic glass, Adv. Mater. 25 (2013) 5904–5908.

[14]  A. Jaiswal, T. Egami, Y. Zhang, Atomic-scale dynamics of a model glass-forming metallic liquid: Dynamical crossover, dynamical decoupling, and dynamical clustering, Phys. Rev. B. 91 (2015) 134204.

[15]  F. Puosi, N. Jakse, A. Pasturel, Dynamical, structural and chemical heterogeneities in a binary metallic glass-forming liquid, J. Phys. Condens. Matter. 30 (2018) 145701.

[16] C.B. Roth, R.R. Baglay, Polymer Glasses, in: C.B. Roth (Ed.), Polymer Glasses, CRC Press, Taylor & Francis Group, 2016.

[17] G.W. Scherer, Editorial comments on a paper by Gordon S. Fulcher, J. Am. Ceram. Soc. 75 (1992) 1060–1062.

[18] A. Jaiswal, T. Egami, K.F. Kelton, K.S. Schweizer, Y. Zhang, Correlation between fragility and the Arrhenius crossover phenomenon in metallic, molecular, and network liquids, Phys. Rev. Lett. 117 (2016) 205701.

[19] Q. Wang, S.T. Zhang, Y. Yang, Y.D. Dong, C.T. Liu, J. Lu, Unusual fast secondary relaxation in metallic glass, Nat. Commun. 6 (2015) 7876.

[20] R. Ni, M.A.C. Stuart, M. Dijkstra, Pushing the glass transition towards random close packing using self-propelled hard spheres, Nat. Commun. 4 (2013) 2704.

[21] X. Monnier, D. Cangialosi, B. Ruta, R. Busch, I. Gallino, Vitrification decoupling from α-relaxation in a metallic glass, Sci. Adv. 6 (2020) eaay1454.

[22] S. Karmakar, C. Dasgupta, S. Sastry, Short-time beta relaxation in glass-forming liquids Is cooperative in nature, Phys. Rev. Lett. 116 (2016) 85701.

[23] H.-B. Yu, R. Richert, K. Samwer, Structural rearrangements governing Johari-Goldstein relaxations in metallic glasses, Sci. Adv. 3 (2017).

[24] V.M. Boucher, D. Cangialosi, H. Yin, A. Schönhals, A. Alegría, J. Colmenero, $T_g$ depression and invariant segmental dynamics in polystyrene thin films, Soft Matter. 8 (2012) 5119–5122.

[25] M.D. Ediger, J.A. Forrest, Dynamics near free surfaces and the glass transition in thin polymer films: A view to the future, Macromolecules. 47 (2014) 471–478.

[26] D. Chatterjee, A. Annamareddy, J. Ketkaew, J. Schroers, D. Morgan, P.M. Voyles, Fast surface dynamics on a metallic glass nanowire, n.d.

[27] Q.L. Bi, Y.J. Lü, W.H. Wang, Multiscale relaxation dynamics in ultrathin metallic glass-forming films, Phys. Rev. Lett. 120 (2018) 155501.

[28] Y.J. Lü, W.H. Wang, Single-particle dynamics near the glass transition of a metallic glass, Phys. Rev. E. 94 (2016) 62611.

[29] S.G. Hao, C.Z. Wang, M.J. Kramer, K.M. Ho, Microscopic origin of slow dynamics at the good glass forming composition range in $Zr_{1-x}Cu_x$ metallic liquids, J. Appl. Phys. 107 (2010) 053511.

[30] C. Tang, P. Harrowell, Chemical ordering and crystal nucleation at the liquid surface: A comparison of $Cu_{50}Zr_{50}$ and $Ni_{50}Al_{50}$ alloys, J. Chem. Phys. 148 (2018) 044509.

[31]    T. Egami, Understanding the properties and structure of metallic glasses at the atomic level, JOM. 62 (2010) 70–75.

[32]    L. Berthier, G. Biroli, Theoretical perspective on the glass transition and amorphous materials, Rev. Mod. Phys. 83 (2011) 587–645.

[33]    L. Berthier, Time and length scales in supercooled liquids, Phys. Rev. E. 69 (2004) 020201 (R).

[34]    D.N. Perera, P. Harrowell, Relaxation dynamics and their spatial distribution in a two-dimensional glass-forming mixture, J. Chem. Phys. 111 (1999) 5441–5454.

[35]    N. Kuon, E. Flenner, G. Szamel, Comparison of single particle dynamics at the center and on the surface of equilibrium glassy films, J. Chem. Phys. 149 (2018) 074501.

[36]    S. Plimpton, Fast parallel algorithms for short-range molecular dynamics, J. Comput. Phys. 117 (1995) 1–19.

[37]    M.I. Mendelev, M.J. Kramer, R.T. Ott, D.J. Sordelet, D. Yagodin, P. Popel, Development of suitable interatomic potentials for simulation of liquid and amorphous Cu–Zr alloys, Philos. Mag. 89 (2009) 967–987.

[38]    D.C. Rapaport, The Art of Molecular Dynamics Simulation, 2nd ed., Cambridge University Press, 2004.

[39]    N. Mattern, A. Schöps, U. Kühn, J. Acker, O. Khvostikova, J. Eckert, Structural behavior of $Cu_xZr_{100-x}$ metallic glass (x=35−70), J. Non. Cryst. Solids. 354 (2008) 1054–1060.

[40]    T. Hecksher, A.I. Nielsen, N.B. Olsen, J.C. Dyre, Little evidence for dynamic divergences in ultraviscous molecular liquids, Nat. Phys. 4 (2008) 737–741.

[41]    F. Faupel, W. Frank, M.-P. Macht, H. Mehrer, V. Naundorf, K. Rätzke, H.R. Schober, S.K. Sharma, H. Teichler, Diffusion in metallic glasses and supercooled melts, Rev. Mod. Phys. 75 (2003) 237–280.

[42]    D. Lee, J.J. Vlassak, Diffusion kinetics in binary CuZr and NiZr alloys in the super-cooled liquid and glass states studied by nanocalorimetry, Scr. Mater. 165 (2019) 73–77.

[43]    E.C. Stelter, D. Lazarus, Diffusion and thermal stability of amorphous copper zirconium, Phys. Rev. B. 36 (1987) 9545–9558.

[44]    S.K. Sharma, F. Faupel, Correlation between effective activation energy and pre-exponential factor for diffusion in bulk metallic glasses, J. Mater. Res. 14 (1999) 3200–3203.

[45]    W. Fernengel, H. Kronmüller, M. Rapp, Y. He, The activation energy of crystallization of amorphous $Fe_{40}Ni_{40}P_{14}B_6$, Appl. Phys. A Solids Surfaces. 28 (1982) 137–144.

[46] U. Geyer, S. Schneider, W.L. Johnson, Y. Qiu, T.A. Tombrello, M.-P. Macht, Atomic diffusion in the supercooled liquid and glassy states of the $Zr_{41.2}Ti_{13.8}Cu_{12.5}Ni_{10}Be_{22.5}$ alloy, Phys. Rev. Lett. 75 (1995) 2364–2367.

[47] F.H. Stillinger, A topographic view of supercooled liquids and glass formation, Science. 267 (1995) 1935–1939.

[48] G. Biroli, J.P. Garrahan, Perspective: The glass transition, J. Chem. Phys. 138 (2013) 12A301.

[49] A. Rahman, Correlations in the motion of atoms in liquid argon, Phys. Rev. 136 (1964) A405–A411.

[50] S.K. Kumar, G. Szamel, J.F. Douglas, Nature of the breakdown in the Stokes-Einstein relationship in a hard sphere fluid, J. Chem. Phys. 124 (2006) 214501.

[51] K. Vollmayr-Lee, R. Bjorkquist, L.M. Chambers, Microscopic picture of aging in $SiO_2$, Phys. Rev. Lett. 110 (2013) 17801.

[52] J.W. Ahn, B. Falahee, C. Del Piccolo, M. Vogel, D. Bingemann, Are rare, long waiting times between rearrangement events responsible for the slowdown of the dynamics at the glass transition?, J. Chem. Phys. 138 (2013) 12A527.

[53] M.P. Ciamarra, R. Pastore, A. Coniglio, Particle jumps in structural glasses, Soft Matter. 12 (2016) 358–366.

[54] R.W. Balluffi, S.M. Allen, W.C. Carter, Kinetics of Materials - Page 158, Wiley, 2005.

[55] J. Helfferich, F. Ziebert, S. Frey, H. Meyer, J. Farago, A. Blumen, J. Baschnagel, Continuous-time random-walk approach to supercooled liquids. I. Different definitions of particle jumps and their consequences, Phys. Rev. E. 89 (2014) 42603.

[56] Y. Fan, T. Iwashita, T. Egami, Energy landscape-driven non-equilibrium evolution of inherent structure in disordered material, Nat. Commun. 8 (2017) 15417.

[57] Salez, Thomas, McGraw, Joshua D., Dalnoki-Veress, Kari, Raphaël, Elie, Forrest, James A., Glass transition at interfaces, Europhys. News. 48 (2017) 24–28.

[58] Y.Q. Cheng, E. Ma, Atomic-level structure and structure–property relationship in metallic glasses, Prog. Mater. Sci. 56 (2011) 379–473.

[59] S.S. Schoenholz, E.D. Cubuk, E. Kaxiras, A.J. Liu, Relationship between local structure and relaxation in out-of-equilibrium glassy systems, Proc. Natl. Acad. Sci. 114 (2017) 263–267.

[60] J. Schroers, Processing of bulk metallic glass, Adv. Mater. 22 (2010) 1566–1597.

[61] P. Zhang, J.J. Maldonis, Z. Liu, J. Schroers, P.M. Voyles, Spatially heterogeneous dynamics

in a metallic glass forming liquid imaged by electron correlation microscopy, Nat. Commun. 9 (2018) 1129.

[62] M.L. Rahaman, L.C. Zhang, H.H. Ruan, Effects of environmental temperature and sliding speed on the tribological behaviour of a Ti-based metallic glass, Intermetallics. 52 (2014) 36–48.

# Supporting Information

# Mechanisms of bulk and surface diffusion in metallic glasses determined from molecular dynamics simulations

Ajay Annamareddy, Paul M. Voyles, John Perepezko, and Dane Morgan

Department of Materials Science and Engineering, University of Wisconsin-Madison, Madison, Wisconsin 53706, USA.

## S1. Atomic dynamics of $Cu_{54}Zr_{46}$

In $Cu_{50}Zr_{50}$ simulations, the less-coordinated Cu segregates to the surface, and the composition of the surface (outer 2.5 Å) is $Cu_{54}Zr_{46}$. To check the dynamics of atoms at this surface composition, MD simulations of $Cu_{54}Zr_{46}$ were conducted. $T_g$ of $Cu_{54}Zr_{46}$ is identified to be 720 K ± 10 K, as shown in Fig. S1(a). Fig. S1(b) shows the diffusion coefficients of Cu and Zr atoms in $Cu_{50}Zr_{50}$ and $Cu_{54}Zr_{46}$ at 700 K. As can be observed, there is only a slight change of less than 5% in the diffusivities of Cu and Zr atoms from $Cu_{50}Zr_{50}$ to $Cu_{54}Zr_{46}$. Assuming the surface behaves similarly to the bulk, the increase in the fraction of Cu atoms at the surface leads to an increase in dynamics that are negligible compared to the observed fast surface dynamics in the paper (with $D_s/D_v$ = 26 at 700 K).

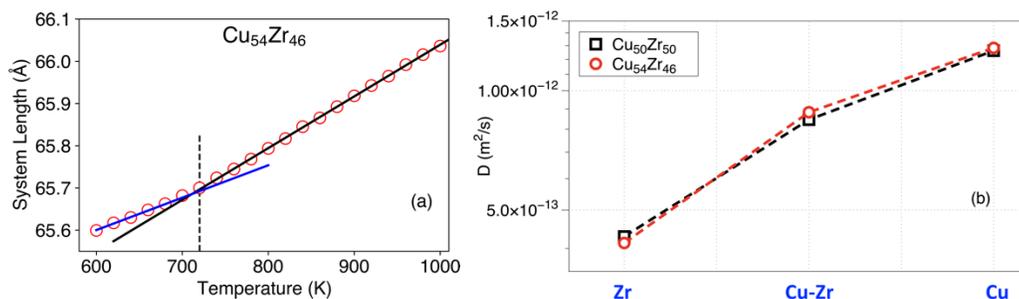

Fig. S1. (a) Identifying $T_g$ in $Cu_{54}Zr_{46}$ from the temperature variation of simulation box length. (b)The ratio of the bulk diffusion coefficients of copper and zirconium in $Cu_{50}Zr_{50}$.

## S2. Relaxation behavior Past the Equilibration Period

When free surfaces are created, atoms in the outer layers have significant strains leading to strain-relaxation of atoms over a period of time. The equilibration period should be sufficiently long to encompass the relaxation effects so that thermally active random diffusion mainly underlies the kinetics thereafter. To demonstrate that relaxation effects past the equilibration are not significant and does not significantly impact kinetics of diffusion, we consider three different ways to assess the changes occurring during the simulation, focusing on both energy and structure.

**Total potential energy:** First, we note that if relaxation effects were driving the atom hopping motion then such relaxation would lead to a decrease in potential energy. In Fig. S2, we check the time evolution of the average potential energy (PE) of all atoms and surface atoms (in the outer 2.5 Å) over a reasonably large time window of 10 ns (following the equilibration period of 1ns) to detect any possible relaxation in $Cu_{50}Zr_{50}$ at 640 K. The variation in the average PE of all the atoms in the slab (non-periodic along an axis) simulation is very small, with a total change of about 2 meV/atom. For surface atoms the average PE changes by about 20 meV/atom (< 0.5%), and actually increases, likely due to variation in Cu segregation to the surface. These small changes in energy for the bulk and surface show that the system is not undergoing any significant relaxation over the timescale of this simulation.

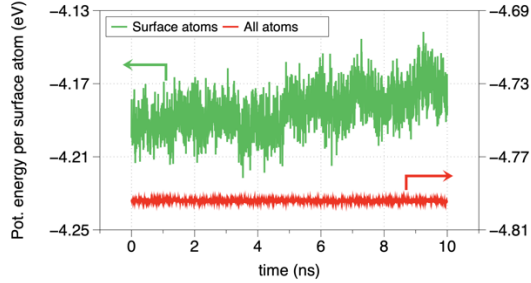

Fig. S2. Plot of average potential energy for surface atoms and all atoms in slab simulation in $Cu_{50}Zr_{50}$ at 640 K.

**Hopping atom potential energy:** Here we continue to explore if atomic motion is reducing PE but focus on individual surface atom hops to avoid missing any effects due to averaging. Specifically, we assess the change in energy of hopping atoms to determine if that change is large enough to represent some form of relaxation-driven hopping occurring past the equilibration period. First, we identify an individual atom as having hopped if it is displaced by more than 1.8 Å during a period of 5 ps. Fig. S3(a) shows the probability distribution function of $\Delta E$ ($\equiv PE_{after-hop} - PE_{before-hop}$) of hopping atoms during 3 different time periods: at the beginning (first 10 ps) and the end (last 100 ps) of the equilibration, and during the ensuing production phase (totaling 10 ns). Relaxation-driven hops can be identified by a lowering of potential energy after the hop, or $\Delta E < 0$. For diffusion hops, with no apparent driving force, $\Delta E$ can be either positive or negative, and should be uniformly distributed around 0. We see from Fig. S3(a) that, as expected, there is appreciable relaxation at the initial stage of the equilibration, with $\Delta E < 0$ hops being dominant and a greater than 100 meV average bias in the hops toward lower energy. However, the $\Delta E$ distributions at the end of the equilibration and during the ensuing production phase are very similar and both exhibit almost symmetric distributions around 0 eV. If relaxation-driven atom hops had persisted through to the end of the equilibration period, we would have seen differences in the distributions in the last 100 ps and production phase. The end of the equilibration and production phase both exhibit a small bias to negative energies. However, the scale of this energy change is ~18 meV/hop on average, which is too small to alter the diffusion coefficient we measure at temperatures above 600 K as it is well below the thermal energy of the system and very far below the activation energies for the surface diffusion process of ~0.95 eV. Even if all 18 meV went directly to lowering the activation energy during hopping it would increase the diffusivity by only about 30% at 640 K.

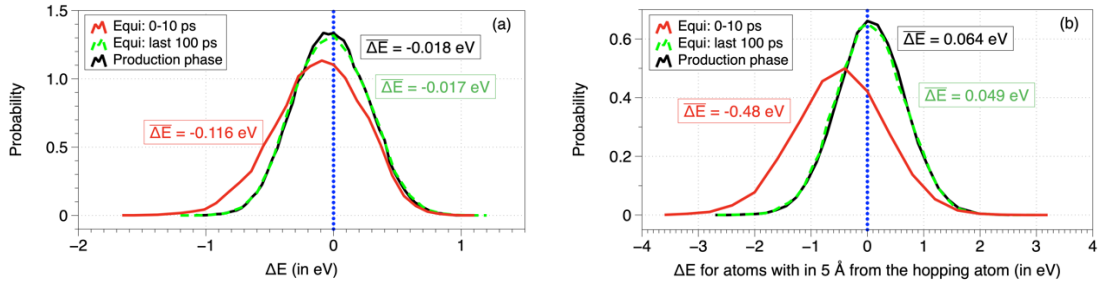

Fig. S3. (a) Probability distribution function of energy difference ($\Delta E$) for surface atom hops during three different time windows: (i) first 10 ps of equilibration period, (ii) last 100 ps of equilibration period, (iii) 10 ns of the ensuing production phase. (b) Similar probability distribution function analysis as in (a) but considering all atoms in a sphere of radius 5 Å from the hopping atom for measuring energy difference.

It is also possible that a relaxation process could lead to decrease in local energy of the system surrounding a hopping atom, and this may not be reflected in the energy change of the individual hopping atom. To verify if such an effect is occurring and has different implications than our single atom analysis, in Fig. S3(b) we show the probability distribution function of energy change of all atoms in a sphere of radius 5 Å from the hopping atom. We see that the end phase of equilibration period has identical distribution to the distribution in the production phase indicating that the 1ns equilibration period applied is sufficiently long enough to accommodate significant relaxation-driven atomic hops. Also, we see that in the equilibration and production phase both exhibit a small bias to negative energies of ~49 meV/hop on average, distributed over many atoms, which is too small to alter the diffusion coefficient as it is still far below the activation energies for the surface diffusion process of ~0.95 eV.

**Voronoi clusters and structural changes:** Here we determined the Voronoi indices of copper atoms in order to study local structures in our surface simulation. While copper atoms forming a perfect icosahedral cluster <0, 0, 12, 0> are the most stable and help in the formation of the glass, in $Cu_{50}Zr_{50}$ only ~6% of Cu atoms have icosahedral structure. As a result, we also consider icosahedral-*like* polyhedra <0, 2, 8, 2>, <0, 2, 8, 1>, <0, 3, 6, 3>, and <0, 1, 10, 2> and call all five polyhedra together the *ico-like* polyhedra [1].

The atoms in the simulation box are divided into different layers, each of width 2.5 Å, as shown in Fig. S4(a) and the concentration of ico-like polyhedra of the layers, averaged over 10 ns, is shown in Fig. S4(b). Very few outer layer Cu atoms (which is the surface in our manuscript) form ico-like polyhedron because of their low coordination number. However, starting from the second layer, the concentration of ico-like clusters increases dramatically compared to the first layer, reaching a value similar to that throughout the slab, and this value is close to the concentration obtained from the bulk simulation (0.29). Also, the concentrations shown in Fig. S4(b) are found to be almost time-invariant. Fig. S4(c) depicts the time variation of concentration of ico-like clusters in the second monolayer illustrating this behavior. This shows that there is no significant structural change of ico-like polyhedra after equilibration for the timescale of our simulations.

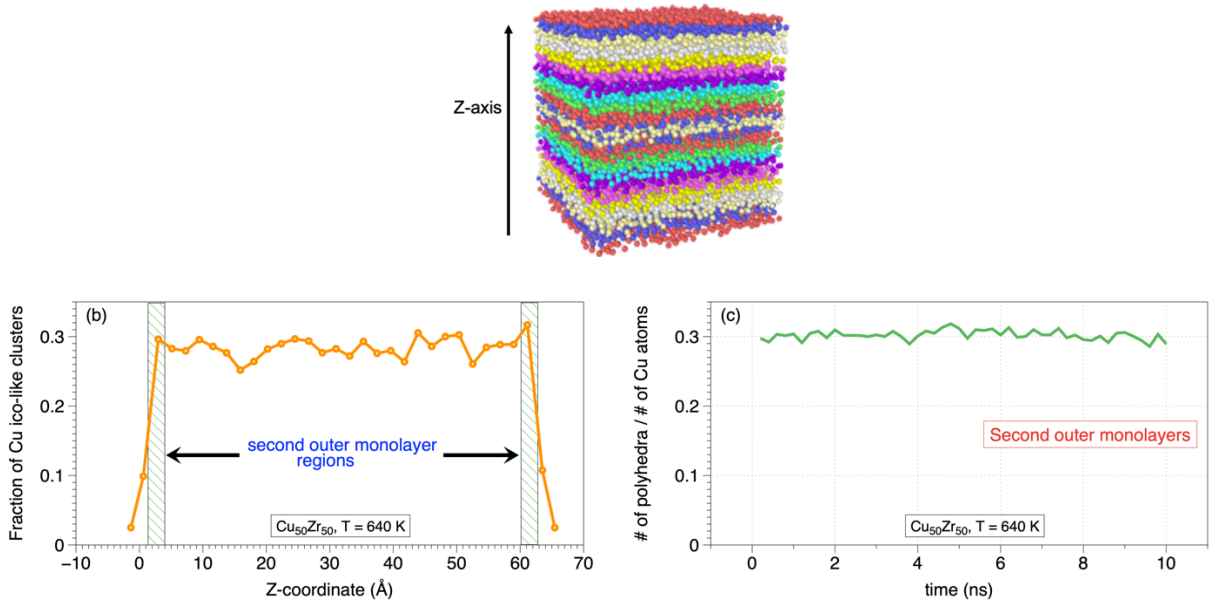

Fig. S4. (a) Dividing simulation box into layers of width 2.5 Å along the direction of non-periodic boundary. (b) Depth variation of the concentration of Cu ico-like clusters in $Cu_{50}Zr_{50}$. (c) Time variation of the fraction of ico-like clusters in the second outer monolayer in $Cu_{50}Zr_{50}$.

## S3. Contrasting dynamics of copper and zirconium atoms in $Cu_{50}Zr_{50}$

Fig. S5(a) shows the bulk diffusion coefficients of Cu and Zr atoms and Fig. S5(b) shows the ratio of the bulk diffusion coefficients of Cu and Zr in $Cu_{50}Zr_{50}$ at different temperatures. The glass transition temperature ($T_g$) is 700 K ± 10 K.

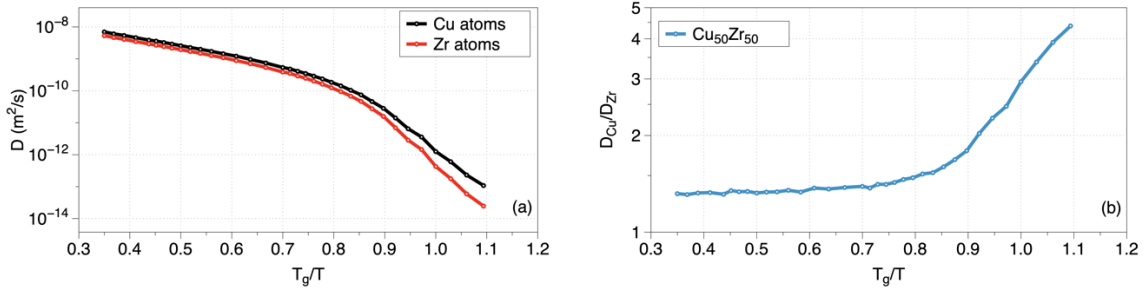

Fig. S5. (a) Variation of bulk diffusion coefficients of Cu ($D_{Cu}$) and Zr ($D_{Zr}$) atoms and (b) $D_{Cu}/D_{Zr}$ with inverse temperature normalized to $T_g$.

## S4. Impact of the choice of surface thickness on different activation energies

In this work, a monolayer of atoms of width 2.5 Å is chosen as representing the surface. In Fig. S6 we show $Q_D$, $Q_1$, and $Q_2$ for surfaces of width 5 Å and 10 Å, in addition to the bulk and surface (2.5 Å width) already shown in Fig. 6 of the manuscript. This analysis serves two purposes: (i) by varying surface thickness, it can be verified if the result in Fig. 6 is independent of the surface thickness chosen. Also, (ii) by extending the surface to 10 Å, the effects of the Cu-rich segregated surface, which occurs mostly within the 2.5 Å width surface, are at least in an average sense largely eliminated as the compositions of the surface and the bulk are almost identical. As expected, with increasing thickness of the surface, all the activation energies in Fig. S6 shift toward the bulk activation energies. However, even for 10 Å width surface layer, considerable difference in $Q_D$ still exists between bulk and surface and this difference mainly arises from the differences in the cage-breaking barrier ($Q_1$).

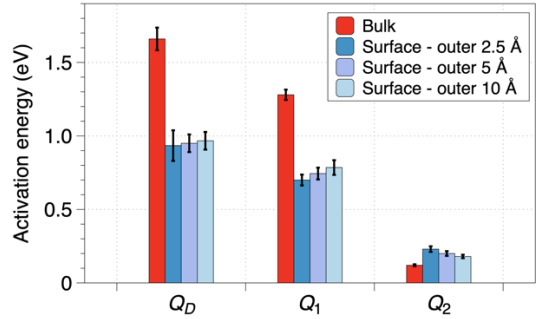

Fig. S6. Comparison of activation energies for total diffusion ($Q_D$), cage-breaking ($Q_1$), and hopping ($Q_2$) for surface and bulk atoms.

## S5. Varying the size of the time-window

Here, we analyze the effect of varying the size of time window on the jump-rate ($\Gamma$) and average of squared jump length ($\langle l^2 \rangle$) of atoms, whose Arrhenius variation determine the cage-breaking barrier ($Q_1$) and hop barrier ($Q_2$), respectively. In general, the size of time window ($\Delta t$) should be larger than average jump period but smaller than a typical caging period. For very high $\Delta t$ (approaching the relaxation time of atoms), multiple jumps of an atom separated by caging events are inappropriately merged into an individual jump. When this merging happens, one identifies too few jumps per unit time and therefore too low a jump rate. As part of the same error, one identifies too large a jump distance for each jump, yielding too large an $\langle l^2 \rangle$. Furthermore, since caging times are lowest at higher temperature, this effect is more pronounced at higher temperature. This temperature dependent lowering of $\Gamma$ reduces the value of $Q_1$ when $\Delta t$ becomes comparable to even the fastest high-temperature caging times in the simulations. A simultaneous increase in $Q_2$ is also expected as the increase in effective squared jump length compared to the true squared jump length is larger at high temperatures. Fig. S7 shows the variation of $Q_1$, $Q_2$, and $Q_D$ (= $Q_1 + Q_2$) for bulk and surface atoms by varying $\Delta t$ around two orders of magnitude from 50 ps to 2 ns.

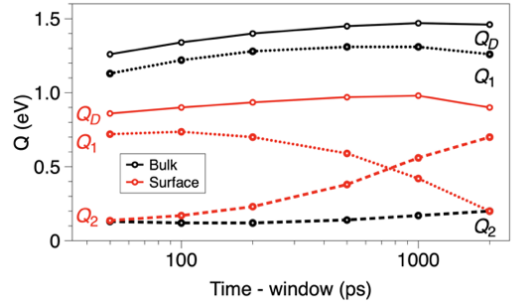

Fig. S7. The effect of varying the size of time-window ($\Delta t$) on individual atom dynamics: Plot of cage-breaking barrier ($Q_1$), hop barrier ($Q_2$) and total activation energy ($Q_D$) for surface and bulk atoms versus $\Delta t$.

The variation of $Q_{D,b}$ and $Q_{D,s}$ is quite small (~10%) over the range of $\Delta t$ studied, as it should be since $Q_D$ approximates the total activation energy for diffusion and is expected to be largely independent of $\Delta t$. However, $Q_1$ for surface atoms begins decreasing for $\Delta t > 100$ ps, which can be understood as emerging from the fact that near 100 ps, $\Delta t$ is approaching the relaxation time of surface atoms at $T_g$. We have chosen $\Delta t = 200$ ps as the longest time we could pick (to assure we almost always capture a full jump) without leading to significant changes in $Q_1$ and $Q_2$. The values of $Q_1$ and $Q_2$ vary by less than 100 meV from 50 ps to 200 ps. The fact that the hop-barriers ($Q_2$) for bulk and surface are almost identical for $\Delta t \leq 200$ ps implies that the main conclusion of our manuscript, that the cage-breaking barrier ($Q_1$) mostly contributes to the differences observed in $Q_D$ between bulk and surface, are valid for any time from 50 ps to 200 ps.

**REFERENCES:**

[1] H. Chen, B. Qu, D. Li, R. Zhou, B. Zhang, Atomic structure and dynamics properties of $Cu_{50}Zr_{50}$ films, J. Appl. Phys. 123 (2018) 025307.